\documentstyle[12pt]{article}
\makeatletter
\@addtoreset{equation}{section}
\def\'{\@ifnextchar'{}{\phantom'}}

\makeatother
\def\beq{\begin{equation}}
\def\eeq{\end{equation}}
\def\beqx{\begin{displaymath}}
\def\eeqx{\end{displaymath}}
\def\beql{\arraycolsep .5mm \begin{eqnarray}}
\def\eeql{\end{eqnarray}}
\def\line{\nonumber \\[2mm] }
\def\lab#1{\label{#1}}
\def\eq#1{(\ref{#1})}

\def\Wir{I}
\def\lag{L}
\def\R{{\sf R}}
\def\C{{\sf C}}
\def\L{{\cal  L}}
\def\H{{\cal  H}}
\def\W{{\cal  C}}
\def\K{{\cal  K}}
\def\D{{\cal  D}}
\def\S{{\cal  S}\'}
\def\bS{\bar\S}
\def\i{{\rm i}}
\def\x{{\rm x}}
\def\y{{\rm y}}
\def\z{{\rm z}}
\def\t{{\rm t}}

\def\del{\partial}
\def\d{{\rm d}}
\def\dd{{\rm D}}
\def\det{{\rm det}}
\def\exp{{\rm exp}}
\def\eins{{\bf 1}}
\def\eps{\varepsilon}
\def\Tr{{\rm Tr}}

\def\O{{\mit \Omega}}
\def\A{A}
\def\F{F}
\def\J{J}
\def\DD{D}
\def\PP{Q}
\def\nabl{\nabla\!}
\def\Pexp{\,{\cal P}{\rm exp}}
\def\sig{\sigma}
\def\gam{\gamma}
\def\bpi{\bar\pi}
\def\bpsi{\bar\psi}
\def\lopar{\lambda}
\def\tsig{\tilde\sig}
\def\loop{\eta}
\def\CC{{\mit\Lambda}}
\def\frac#1#2{{{#1}\over{#2}}}
\def\ft#1#2{{\textstyle{{#1}\over{#2}}}}
\def\^{{}^}
\def\_{{}\!_}
\def\cc{\makebox[0pt][l]{${}^*$}{}}
\def\E#1#2{E_{#1}\^{#2}}
\def\e#1#2{e_{#1}\^{#2}}
\def\ee#1#2{{\tilde e}_{#1}\^{#2}}
\def\intd#1#2{\int {\rm d}^{#1}\! #2 \,}
\def\op#1{\widehat{#1}}
\def\grp#1({{\rm #1}(}
\def\alg#1({{\bf #1}(}
\def\pois#1#2{\big\{ #1 , #2  \big\}}
\def\komm#1#2{\big\lbrack #1 , #2 \big\rbrack }
\def\deldel#1/#2/{\frac{\del #1}{\del #2}}
\def\deltadelta#1/#2/{\frac{\delta #1}{\delta #2}}

\def\WDW{Wheeler DeWitt }

\begin{document}
\title{About Loop States in Supergravity}
\author{Hans-J\"urgen Matschull\thanks{Work supported by DFG}\\
            II. Institut f\"ur Theoretische Physik\\
            Universit\"at Hamburg\\
            Luruper Chaussee 149\\
            22761 Hamburg\\
            Germany}
\date{March 11, 1994}
\let\nnewpage=\newpage
\let\newpage=\relax
\begin{flushright}
gr-qc/9403034\\
DESY 94-037
\end{flushright}
\maketitle
\let\newpage=\nnewpage
\begin{abstract}
The Wilson loop functionals in terms of Ashtekar's variables were
the first (formal) solutions to the quantized hamiltonian constraint
of canonical gravity.
Here it is shown that the same functionals also solve the supergravity
constraints and some evidence is presented that they are artificially
generated by multiplying the constraints by the metric determinant,
which has become a widely accepted procedure.
Using the same method in 2+1 dimensional gravity and supergravity leads
to wrong results, e.g.~2+1 gravity is no longer a purely topological
theory.
As another feature of the densitized constraints it turns out that the
classical theory desribed by them is not invariant under space
time diffeomorphisms.
\end{abstract}

In the main part of this paper we will focus on the comparison
of the metric and connection representation of supergravity.
Both will be derived using Ashtekar's variables
\cite{ashtekar:87}, which simplify the construction of the
canonical theory, even in the metric formalism, considerably.
For the N=1 theory in 3+1 dimensions the canonical treatment
`a la Dirac' \cite{dirac:65} in the metric (or vierbein)
representation leads to first class constraints which in the
quantized version become inhomogeneous second order
differential equations \cite{eath:84} with non-analytic
coefficients, whereas in the
connection representation they are polynomials in the canonically
conjugate variables and homogeneous in
the momentum variables \cite{jacobson:88}.

But, as for the connection representation of pure gravity,
one has to multiply the supersymmetry constraint by the determinant
of the spatial metric to obtain a polynomial expression.
However, in supergravity this multiplication does not only
produce a `densitized' \WDW operator but also the diffeomorphism
constraint becomes a density of weight 1, because both are
generated by the commutator of the supersymmetry constraint with its
conjugate.

We will see that the densitized supersymmetry constraints
are formally solved by the {\em same} Wilson loop functionals
which are known to solve the \WDW constraint of pure gravity
\cite{jacobson.smolin:88}. In contrast to pure gravity, however,
a single loop functional
now solves {\em all} constraints without considering
functionals that depend on knot classes only, i.e.~the
state functionals are no longer invariant under spatial
diffeomorphism.

Though this result comes out when dealing with supergravity,
it can be reproduced for pure gravity, too, just by dropping all the
fermionic quantities from the action and the constraints.
When defining the polynomial constraints in this way, a single loop
functional again becomes a solution to {\em all} constraints.
Obviously, multiplying the diffeomorphism constraint with
extra vierbein factors destroys the invariance of the theory under
spatial coordinate transformations.
In fact, for the classical theory we will explicitly see that
the invariance under space time diffeomorphisms is lost, if
we use the densitized hamiltonian constraint {\em and} allow the
metric to be singular.

Another property of the loop functionals in supergravity is that
they are purely bosonic states, i.e.~they do not depend on the
fermionic variables. There has been a discussion about the existence
of such states \cite{eath:93,dewit.matschull.nicolai:93,page:93,%
freedman.et.al:94}.
For the metric representation it was shown in \cite{page:93} that
purely bosonic states do not exist, because they cannot fulfill
one of the supersymmetry constraints.
This suggests that
the Wilson loop states are nothing but
solutions artificially generated by multiplying the constraints
with the metric determinant. Note that the solutions themselves
are annihilated by this determinant.

To make all these arguments more precise, in the second part we will
discuss N=2 supergravity in 2+1 dimensions, which is completely
soluble both on the classical and the quantum level.
It will be
shown that the metric and connection representation are
equivalent. However, to define the connection representation
properly, no extra factors of the dreibein are needed to make the
constraints polynomial, and both the metric and the connection
representation are able to deal with singular metrics.

But, if one starts from the four dimensional densitized
constraints and reduces them to three dimensions, or, equivalently,
if one multiplies the hamiltonian constraint by the metric determinant
in the same way as for the 3+1 dimensional theory, one obtains
a different set of constraints, the `densitized connection
representation'. In this representation, one again finds the
loop states and the quantum theory becomes completely
different. In particular, there are now infinitly many states
even if the spatial
topology is trivial and one requires the state to be invariant
under diffeomorphisms (which does not follow from the constraints!),
whereas in the `correct' quantum theory there are only
finitely many classical degrees of freedom, i.e~the wave
functional depends on finitely many variables only, and there is only
one state for trivial topology.

\section{N=1 supergravity in four dimensions}

The canonical formalism for N=1 supergravity in terms of
Ashtekar's new variables has been worked out in
\cite{jacobson:88}. We will use slightly different notation
here to make the results formally as similar as possible to the
corresponding results for the three dimensional theory below.

The first order action for N=1 supergravity is usually written as
\cite{deser.zumino:76,ferrara.freedman.nieuwenhuizen:76}
\beq
  \Wir'[E,\O,\psi,\bpsi] =
  \intd4x   \lag'_{\rm EH} + \lag'_{\rm RS} ,
\eeq
where
\beql
  \lag'_{\rm EH} &=& \ft12 E \E AM \E BN R_{MN}\^{AB}[\O],\line
  \lag'_{\rm RS} &=& \i \eps^{MNPQ} \big(
     \bpsi_M \sig_N D_P \psi_Q
   - D_M\bpsi_N \sig_P \psi_Q \big)
\eeql
are the Einstein Hilbert and Rarita Schwinger action, respectively.
The notation is as follows.
Indices from the beginning of the alphabet always denote flat
tangent space vectors, those from the middle of the alphabet
curved space time indices. The vierbein $\E AM$ (or its
inverse $\E MA$) thus has
the flat index $A$ running from $0$ to $3$, raised and lowered
by the lorentzian metric $\eta_{AB}={\rm diag}(-,+,+,+)$,
and the curved index $M$, taking the values $\t,\x,\y,\z$.
$E$ is the determinant of $\E MA$ and $\eps^{MNPQ}$ the
Levi Civita tensor density with $\eps^{\t\x\y\z}=1$.

The spin connection $\O_{MAB}$ defines the covariant derivative
of a tangent space vector with flat index
\beq
    D_M V_A = \del_M V_A + \O_M\^A\_B V^B  .
\eeq
The field strength or curvature of $\O$ is given by
\beq
  R_{MNAB}[\O] = \del_M \O_{NAB}  - \del_N \O_{MAB}
           + \O_{MA}\^C \O_{NCB} - \O_{NA}\^C \O_{MCB}.
\eeq

We introduce Ashtekar's variables by mapping the $4\cdot6$ components
of the real $\alg so(3,1)$
            spin connection $\O_{MAB}$ onto $4\cdot3$ complex
components of the $\alg so(3,\C)$ connection $\A_{Ma}$, $a=1,2,3$.
The mapping is given by
\beq
  \A_{Ma} = \J_{aAB} \O_M\^{AB}  , \ \ \ \
  \A\cc_{Ma} = \J\cc_{aAB} \O_M\^{AB},
\lab{A=JO}
\eeq
where the coefficients $\J_{aAB}$ and $\J\cc_{aAB}$ form a basis
of selfdual and antiselfdual antisymmetric tensors, i.e.~%
\beq
  \J_{aAB} = -\ft\i2 \eps_{AB}\^{CD} \J_{aCD},  \ \  \ \
  \J\cc_{aAB} =  \ft\i2 \eps_{AB}\^{CD} \J\cc_{aCD},
\lab{J=epsJ}
\eeq
where $\eps^{ABCD}$ is the flat Levi Civita tensor with
$\eps^{0123}=-\eps_{0123}=1$.
The $\J$-symbols are complete and orthonormal
\beqx
  \J_{aAB} \J_a\^{CD} + \J\cc_{aAB} \J\cc_a\^{CD} =
  \delta_{[A}^{[C} \delta_{B]}^{D]}
       = \ft12 \delta_A^C \delta_B^D - \ft12 \delta_A^D \delta_B^C,
\eeqx
\beq
  \J_{aAB} \J_b\^{AB} = \J\cc_{aAB} \J\cc_b\^{AB} =
                        \eta_{ab} , \ \ \ \
  \J_{aAB} \J\cc_b\^{AB} = 0 ,
\lab{J-prop}
\eeq
where $\eta_{ab}=\delta_{ab}$ is the `spatial' component of the
flat metric (remember that $a=1,2,3$ whereas $A=0,1,2,3$).
Thus the relation \eq{A=JO} can be inverted to give
\beq
  \O_{MAB} = \J_{aAB} \A_{Ma} + \J\cc_{aAB} \A\cc_{Ma}.
\eeq

An explicit representation for $\J$ is
\beq
  \J_{aAB} =  \ft\i2 \eta_{aA} \delta^0_B - \ft\i2 \eta_{aB} \delta^0_A
       - \ft12 \eps^0\_{aAB}.
\eeq
Here we used that the range of the small indices is just a subset
of that of the big indices.
Another useful property of $\J$ is that it provides a four dimensional
representation of $\alg so(3)$, which commutes with its conjugate
\beql
  \J_{aA}\^B \J_{bB}\^C &=& -\ft14 \eta_{ab} \delta_A^C
            + \ft12 \eps_{abc} \J_{cA}\^C ,  \line
  \J\cc_{aA}\^B \J\cc_{bB}\^C &=& -\ft14 \eta_{ab} \delta_A^C
            + \ft12 \eps_{abc} \J\cc_{cA}\^C ,  \line
  \J\cc_{aA}\^B \J_{bB}\^C &=&
  \J_{bA}\^B \J\cc_{aB}\^C,
\lab{J-so(3)}
\eeql

Thus $\A_{Ma}$ is a $\alg so(3,\C)$ connection acting on
selfdual tensors $T_{AB}=\J_{aAB}T_a$ as
\beq
   D_M T_a = \del_M T_a + \eps_{abc} \A_{Mb} T_c
\eeq
and its field strength is given by
\beq
   \F_{MNa}[\A] = \del_M \A_{Na} - \del_N \A_{Ma}
             + \eps_{abc} \A_{Mb} \A_{Nc}.
\eeq
It follows from \eq{A=JO} that this field strength is related to
the curvature $R_{MNAB}$ by
\beq
   \F_{MNa} = \J_{aAB} R_{MN}\^{AB} , \ \ \ \
   \F\cc_{MNa} = \J\cc_{aAB} R_{MN}\^{AB} ,
\eeq
or
\beq
  R_{MNAB} = \J_{aAB} \F_{MNa} + \J\cc_{aAB} \F\cc_{MNa}.
\eeq

With the last formula the Einstein-Hilbert action
splits into a part holomorphic in $\A$ and its conjugate.
So after some algebra, which uses the selfduality of $\J$,
   $\lag'_{\rm EH}$ becomes the real part of
\beq
  \lag_{\rm EH} = -\ft\i2 \eps^{MNPQ} \, \E MA \E NB \,
        \J_{aAB} \F_{PQa}[\A].
\eeq

As in \cite{jacobson:88} we represent the gravitinos as
complex 2-component Gra\ss mann valued spinors, transforming in the
selfdual complexified
$\alg su(2)$ representation of the Lorentz group.
But we will not write out the spinor indices explicitly. Instead,
we introduce the four hermitian two by two matrices $\sig_A$
with
\beq
  \sig_0 = \pmatrix{ 1 & 0 \cr 0 & 1 \cr }, \ \ \ \
  \sig_1 = \pmatrix{ 0 & 1 \cr 1 & 0 \cr }, \ \ \ \
  \sig_2 = \pmatrix{ 0 & -\i \cr \i & 0 \cr }, \ \ \ \
  \sig_3 = \pmatrix{ 1 & 0 \cr 0 & -1 \cr }.
\eeq
Matrices with curved indices are obtained by multiplication
with a vierbein: $\sig_M = \E MA \sig_A$.
Note that the algebra of these matrices is given by the
$\J$-symbols
\beq
  \sig_A \sig_a = 2\i \J_{aAB} \sig^B , \ \ \ \
  \sig_a \sig_A =-2\i \J\cc_{aAB} \sig^B.
\eeq

The covariant derivative of a spinor $\chi$ then reads
\beq
    D_M \chi = \del_M \chi - \ft\i2 \A_{Ma} \, \sig_a \chi.
\eeq
A Dirac conjugate spinor, which is defined
by $\bar\chi=\chi^\dagger(\i\sig_0)=\i\chi^\dagger$, transforms under
the complex conjugate or antiselfdual
representation and its covariant derivative
is
\beq
   D_M \bar\chi = \del_M \bar\chi + \ft\i2 \A\cc_{Ma}
                   \, \bar\chi \sig_a .
\eeq
Using the properties of $\J$ it is straightforward to show
that $\i\bar\chi\sig_A\chi$ is real and
                          transforms as a vector.
Note that $\bar\chi\chi$ is not a scalar but the $0$-component
of a vector.

With these definitions we see immediately that also the
Rarita Schwinger action splits into a part holomorphic in
$\A$ and its conjugate, thus $\lag'_{\rm RS}$ is the real part of
\beq
   \lag_{\rm RS} =  2\i \eps^{MNPQ} \,
     \bpsi_M \sig_N D_P \psi_Q .
\eeq{}

The total action $\Wir'$ is given as the real part of
$\Wir[E,\A,\psi,\bpsi]=\intd4x\lag_{\rm EH}+\lag_{\rm RS}$.
As this is a holomorphic function of $\A$, the
equations of motion for $\A$ are the same as those for
$\O$ in $\Wir'[E,\O,\psi,\bpsi]$, which directly follows
from the Cauchy Riemann differential equations for holomorphic
functions.
As shown in \cite{jacobson:88}, the complex action becomes real
(up to a total derivative) if the equations of motion for
$\A_{Ma}$ are satisfied and thus $\Wir$ and $\Wir'$
imply the same equation for all fields.
Therefore we can take the complex action as starting point for
the canonical quantization.

\subsection*{The metric representation}

Though our lagrangian
\beq
  \lag =    \i\eps^{MNPQ} \big(
    -\ft12 \E MA \E NB \J_{aAB} \F_{PQa}
    + 2 \, \bpsi_M \sig_N D_P \psi_Q \big)
\lab{lag-asht}
\eeq
is written in terms of Ashtekar's variables, it is still possible
to reproduce the metric (or vierbein) representation of D'Eath
\cite{eath:84}. With the help of Ashtekar's variables and the
complex action $\lag$, however, the derivation of the
constraint operators becomes considerably simpler. Thus we
will show this construction briefly.

As we shall not use the variables $\A_{Ma}$ as canonical variables,
we use 1.5 order formalism here, i.e.~$\A_{Ma}$ is a
function of the other fields $\E MA$ and $\psi_M$, which
is defined by its equations of motion.
We already know that these equations are the same as
those for the real lagrangian $\lag'$. The solutions are
\beq
  \A_{Ma}[E,\psi] = \J_{aAB} \O_M\^{AB},
\eeq
where $\O_{MAB}$ is implicitly given by
\beq
   \del_{[M} \E {N]}A + \O_{[M}\^A\_B \E {N]}B
     - \bpsi_{[M} \sig^A \psi_{N]}=0 .
\eeq
The last equation is just the usual torsion equation
obtained by differentiating $\lag'$ with respect
to $\O_{MAB}$.

We now split the space time spanned by the coordinates $M=\t,\x,\y,\z$
into a spatial hypersurface spanned by $m=\x,\y,\z$ and a
coordinate $\t$ which serves as the canonical time variable and
which we assume to be a globally defined coordinate.
By inserting this into the lagrangians we obtain
\beql
  \lag_{\rm EH} &=&
    - \i \eps^{mnp} \E mA \E nB \J_{aAB} F_{\t pa}
    - \i \eps^{mnp} \E \t A \E mB  \J_{aAB} F_{npa} \line
 &=& 2 \i \eps^{mnp} \del_\t \E mA \, \E nB \J_{aAB} \A_{pa}\line
 &&{} - \i \eps^{mnp} D_p \big( \E mA \E nB \J_{aAB}\big)\, \A_{\t a}
    - \i \eps^{mnp} \E \t A \E mB  \J_{aAB} F_{npa} \line
  \lag_{\rm RS} & = &
  2\i \eps^{mnp} \big (
     \bpsi_\t \sig_m D_n \psi_p
   - \bpsi_m \sig_\t D_n \psi_p  \line & &  \hspace*{6em}   {}
   + \bpsi_m \sig_n D_\t \psi_p
   - \bpsi_m \sig_n D_p \psi_\t \big).
\eeql

Here we have integrated by parts the Einstein Hilbert term
to get a lagrangian without second order time derivatives.
Note that $\A_{Ma}$ is a function of $\E MA$ and $\psi_M$
and the derivatives of $\E MA$.
As already mentioned, the imaginary
part of $\lag$ is a total derivative, thus in fact we have
a real action. A general discussion of the canonical
formalism for a lagrangian of this type is given in
\cite{fukuyama.kamimura:90}.

We can read off the momenta of $\E mA$ and $\psi_m$. They are
\beql
   P_A\^m &=&\deltadelta \lag / (\del_\t \E mA) / =
     2\i \eps^{mnp} \J_{aAB} \E nB  \A_{pa}, \line
   \bpi^m &=& \deltadelta \lag / (\del_\t \psi_m) / =
     -2\i \eps^{mnp} \, \bpsi_n \sig_p .
\lab{metric-momenta}
\eeql

Spinor derivatives always act from the left, which produces the
extra sign in the equation for $\bpi^m$.
As there are no `mixing momenta', i.e.~the canonical variables
split into configuration variables $\E mA$ and $\psi_m$ and
momentum variables $P_A\^m$ and $\bpi^m$, there is no need to
compute dirac brackets. Instead, we can read off the correct brackets
directly from
\beq
   \pois{ \E mA} { P_B\^n } = \delta^A_B \delta^n_m , \ \ \ \
   \pois{ \psi_m } {\bpi^n} = - \delta^n_m \eins .
\eeq
Note the sign of the fermionic Poisson bracket, which has to be
chosen such that the brackets reproduce the correct equations of motion.
$\eins$ is the two by two unity matrix.

The constraints are obtained directly from $\lag$
by differentiating with respect to the Lagrange multipliers
$\E \t A$, $\psi_\t$ and $\bpsi_\t$.
They are
\beql
  \H_A &=& - \i \eps^{mnp} \E mB \J_{aAB} \F_{npa}
            - 2\i \eps^{mnp} \, \bpsi_m \sig_A D_n \psi_p , \line
  \S & = &  2\i \eps^{mnp} \sig_m D_n \psi_p , \line
  \bS & = &  2\i \eps^{mnp} D_m ( \bpsi_n \sig_p ) ,
\eeql
where $\H_A$ is the combined \WDW constraint together with
the generators of spatial diffeomorphism. This combination is
useful in supergravity because it is just the commutator of
the supersymmetry constraints $\S$ and $\bS$.

As $\A_{ma}$ appears here, they are rather complicated functions
of the canonical fields. Note, however, that all Lorentz covariant
derivatives contain $\A_{ma}$ and not its conjugate, because
$\psi_m$ as well as $\bpsi_m\sig_n$ transform under the
selfdual representation of the Lorentz group.

                         An additional constraint follows from
the definition of $P_m\^A$ as a function of the velocities
$\del_t \E mA$ (which are implicit in $\A_{pa}$ in \eq{metric-momenta}).
Of course, this constraint is just the equation of motion
for $\A_{\t a}$, i.e.~the $\t$-component of the torsion equation.
The derivative of $\lag$ with respect to $\A_{\t a}$ is
\beq
   - \i \eps^{mnp} D_m \big(\E nA \E pB \J_{aAB} \big)
   + \eps^{mnp}\,  \bpsi_m \sig_n \sig_a \psi_p   =0
\eeq
Writing out the covariant derivative explicitly and using
\eq{J-so(3)} we get the Lorentz constraint
\beq
  \L_a = - \i \eps^{mnp} \del_m \big( \E nA \E pB \J_{aAB} \big)
          -  \J_{aA}\^B \E mA P_B\^m
   + \ft\i2  \bpi^m \sig_a \psi_p
\eeq

To simplify the notation, we now introduce a
`selfdual densitized dreibein' defined by
\beq
  \ee ap = -\eps^{mnp} \E mA \E nB \J_{aAB}.
\lab{ee-def}
\eeq
Note that this is complex and obeys certain reality conditions
which we are not interested in.
Its determinant $\tilde e=\det\ee am$ is given by the
determinant of the three metric $g_{mn}=\E mA E_{nA}$, thus
$\ee am$ can be inverted and divided by the determinant to give
an inverse dreibein $e_{am}$ which defines the spatial
metric via $g_{mn}=e_{ma}e_{na}$.
A straightforward calculation shows that these $g_{mn}$ are in fact
the spatial components of the four metric\footnote
{If one uses a gauge fixed vierbein with $\E m0=0$, i.e.~the
timelike unit covector pointing along the $\t$ axis, the
selfdual dreibein becomes real and equal to the spacelike
part of the vierbein.}
$G_{mn}=E_{mA} \E nA$.
The dreibein also defines a spatial $\alg so(3,\C)$ spin connection
$\omega_{ma}$ by
\beq
   \del_{[m} e_{n]a} + \eps_{abc} \omega_{[mb} e_{n]c} =
   \nabl_{[m} e_{n]a} = 0,
\eeq
where $\nabl$ denotes the full covariant derivative on the
spatial hypersurface with respect to the dreibein $e_{ma}$.
As all derivatives appearing here are antisymmetric the
Christoffel connection can be omitted.
We can also define the full covariant derivative
of the spinor $\psi_n$ and the momentum $\bpi^m$, both
transforming in the selfdual representation.
The combinations not including the Christoffel connection
are (note that $\bpi^m$ is a density of weight 1)
\beq
   \nabl_{[m} \psi_{n]} = \del_{[m} \psi_{n]}
                    - \ft\i2 \omega_{[ma} \, \sig_a \psi_{n]}, \ \ \ \
   \nabl_m \bpi^m = \del_m \bpi^m
            + \ft\i2 \omega_{ma} \, \bpi^m \sig_a.
\eeq

Using this the Lorentz constraint can be written as
\beq
  \L_a =  \i \del_m \ee am
          -  \J_{aA}\^B \E mA P_B\^m
   + \ft\i2\,  \bpi^m \sig_a \psi_m
\eeq

We now want to quantize the theory in the $E$-$\psi$-representation.
The wave functional $\Psi$ thus depends on $\E mA$ and
$\psi_m$, and the operators have to satisfy
\beq
   \komm { \op{P_B\^n} }{ \op{\E mA}} =\i \hbar
                                       \delta^A_B \delta^n_m , \ \ \ \
   \komm{\op{\psi_m} } {\op{\bpi^n}} = \i \hbar
                                           \delta^n_m \eins .
\eeq

The simplest choice for $\op P$ and $\op{\bpi}$ would be
$\i\hbar\delta/\delta E$ and $\i\hbar\delta/\delta\psi$, respectively.
But because of the $\del_m\ee am$ term in $\L_a$, which
then becomes a multiplicator, the Lorentz constraint would fail
to generate proper Lorentz transformations on the wave
functional. To avoid this, one has to start from a more
general representation given by
\beq
  \op{P_A\^m} = \i\hbar\deltadelta / \E mA /
              -\i\deltadelta G[E] / \E mA /,
  \ \ \ \
  \op{\bpi_{m\alpha}} = \i\hbar\deltadelta / \psi_{m\alpha} / ,
\eeq
where $G[E]$ is a functional of the fields $\E mA$ and
$\alpha=1,2$ denote the spinor index.
We will write out the spinor index only if its contraction
is not simply given by matrix multiplication.
With this representation the quantized Lorentz constraint reads
\beq
  \op{\L_a} =  \i \del_m \ee am
          -\i \J_{aA}\^B \E mA \Big( \hbar
         \deltadelta / \E mB /  -  \deltadelta G[E] / \E mB / \Big)
   + \ft12\hbar
          \,  ( \sig_a \psi_m )_\alpha\, \deltadelta / \psi_{m\alpha}/.
\eeq
Now assume that $G$ is given as a function of the
$\ee ma$. Then we have
\beql
   \i\J_{aA}\^B \E mA \, \deltadelta G / \E mB /
  &=&-2\i \eps^{mnp} \J_{aA}\^B \J_{bBC}  \E mA  \E nC
                         \deltadelta G / \ee bp /    \line
  &=&- \i \eps^{mnp} \eps_{abc} \J_{cAC} \E mA \E nC
                         \deltadelta G / \ee bp /    \line
  &=&    \i \eps_{abc} \ee cm \deltadelta G / \ee bm / .
\eeql
This obviously cancels against the $\del_m\ee am$ term if
the derivative of $G$ gives the spatial spin connection
$\omega_{mb}$. Such a functional, however, is well known and serves
as a generating functional for the canonical transformation
from the metric to the connection representation
\cite{henneaux.schomblond.nelson:89,matschull.nicolai:92}.
The functional has the simple form
\beq
  G[E] = \ft12 \intd3x \eps^{mnp} e_{ma} \del_n e_{pa} ,
\eeq
however, the dependence of $G$ on $\ee am$ and thus $\E mA$
is highly nonlinear because of the inverse dreibein $e_{ma}$ in
its definition.

The explicit operator for $P_A\^m$ is now given by
\beq
  \op{P_A\^m} =
    \i\hbar \,  \deltadelta / \E mA /
   +2\i \eps^{mnp} \J_{aAB} \E nB \omega_{pa}.
\eeq
It is therefore useful to introduce the quantity
\beq
   \PP_A\^m = 2\i\eps^{mnp} \J_{aAB} \E nB (\A_{pa}-\omega_{pa}),
\eeq
which is represented by the quantum operator
\beq
   \op{\PP_a\^m} = \i\hbar \deltadelta / \E mA / .
\eeq

The Lorentz constraint now generates (selfdual) Lorentz
transformations on the wave functional:
\beq
  \op{\L_a} =- \i \hbar \J_{aA}\^B \E mA \, \deltadelta / \E mB /
            + \ft12\hbar \, (\sig_a \psi_m)_\alpha
                 \, \deltadelta/\psi_{m\alpha} /.
\eeq
To obtain a real constraint $\L_A\^B$ in the $\grp SO(3,1)$
representation of the Lorentz group, one has to compute the
real part of $\L_a\J_{aA}\^B$. As $P_A\^m$ itself
is not real (because of the complex lagrangian), this becomes
a more involved but straightforward calculation. The result is, as
expected,
\beq
  \op{\L_A\^B}= -\i\hbar  \E m{[B}  \deltadelta  / \E m{A]} /
            + \ft12 \J_{aA}\^B \,\hbar \,  (\sig_a \psi_m)_\alpha
                 \, \deltadelta / \psi_{m\alpha} / .
\eeq

The supersymmetry constraints of \cite{eath:84} are now
easily reproduced.
For $\S$ we get
\beql
    \S   & = &  2\i \eps^{mnp}\,  \sig_m D_n \psi_p
     \line & = & 2 \i \eps^{mnp} \, \sig_m \nabl_n \psi_p
          + \eps^{mnp} \, \sig_m \sig_a\psi_p\, (\A_{na}-\omega_{na})
     \line & = & 2 \i \eps^{mnp} \, \sig_m \nabl_n \psi_p
  + 2\i \eps^{mnp} \E mA \J_{aAB} \, \sig^B \psi_p
              \,(\A_{na}-\omega_{na})
     \line & = & 2 \i \eps^{mnp} \, \sig_m \nabl_n \psi_p
          - \sig^A \psi_m  \, \PP_A\^m .
\eeql
and the same calculation for $\bS$ yields
\beql
    \bS   & = &  2\i \eps^{mnp}\,   D_m(\bpsi_n \sig_p )
   \phantom{   2\i \eps^{mnp} \E mA \J_{aAB} \, \sig^B \psi_p
              \,(\A_{na}-\omega_{na}) }
     \line & = &  - \nabl_m \bpi^m
        + \bpsi_m \sig^A \, \PP_A\^m.
\eeql
The quantum operator for $\S$ is obtained simply by inserting
the operator for $\PP_A\^m$:
\beq
   \op{\S} =
   2\i \eps^{mnp} \, \sig_m \nabl_n \psi_p
     - \i\hbar \sig^A \psi_m \,  \deltadelta / \E mA / ,
\eeq
which is exactly the same as the operator given in equation
(4.6) of \cite{eath:84}.
To write down the operator for
$\bS$ we have to give an explicit representation for
$\bpsi_m$, because $\bS$ does not depend on $\bpsi_m$
via $\bpi^m$ only. We have to invert the relation
\eq{metric-momenta}, thus we have to find a matrix $\DD_{mq}$ with
\beq
        \eps^{mnp}\, \sig_p
             \DD_{mq}  = \delta^n_q  \eins,
\eeq
which then gives us (see \cite{eath:84} for an explicit expression for
$\DD_{mn}$)
\beq
   \bpsi_m = \ft\i2 \bpi^n \DD_{nm} , \ \ \ \
  \op{{\bpsi}}_m = -\ft12  \hbar
          \deltadelta / \psi_n / \, D_{nm}.
\eeq

Inserting this we can also reproduce the conjugate constraint
(equation (4.10) of \cite{eath:84}), which reads
\beq
   \op{\bS} =
     - \i \hbar \nabl_m \Big( \, \deltadelta / \psi_m / \, \Big)
     - \ft\i2 \hbar^2   \,  \deltadelta /  \psi_n / \, D_{nm} \sig^A
     \, \deltadelta / \E nA   / .
\eeq

The representations of the remaining constraints $\H_A$ are rather
cumbersome and we will not give them here.
$\H_A\sig^A$ is the bracket of $\S$ with $\bS$. Thus every solution
to the supersymmetry and Lorentz constraint is always a solution
to all constraints, if we define the operator ordering for
$\H_A$ by
\beq
   \op{\H_A} \sig^A = \komm{\op{\S}}{\op{\bS}}
\eeq

There are some ans\"atze for solutions of the supersymmetry constraints
in this representation \cite{eath:93,%
dewit.matschull.nicolai:93,page:93,freedman.et.al:94},
but so far no exact solution is known. See, however, \cite{page:93},
where it is shown that
the constraints cannot be solved by a purely
bosonic state, i.e.~a wave functional that does not depend on the
fermionic variables.

It is reasonable that there
are no purely bosonic states, because the
wave functional should be invariant under supersymmetry transformations
and thus it cannot depend on the bosonic configuration only, as long
as there are no bosonic fields that are invariant under supersymmetry.
For the densitized constraints in the connection representation,
however, we will find purely bosonic solutions and this
may be interpreted as a first hint that there is `something
wrong' with the loop states.

\subsection*{The connection representation}
Again we start with the lagrangian
\beq
  \lag =    \i\eps^{MNPQ} \big(
    -\ft12 \E MA \E NB \J_{aAB} \F_{PQa}
    + 2 \, \bpsi_M \sig_N D_P \psi_Q \big),
\eeq
but now we will use first order formalism. Thus the independent
variables are $\E MA$, $\A_{Ma}$, $\bpsi_M$ and $\psi_M$.
The momenta of $\A_{ma}$ and $\psi_m$ are
\beql
    \deltadelta \lag / (\del_\t \A_{ma}) / & = &
      -\i\eps^{mnp} \E nA \E pB \J_{aAB}   = \i \ee am , \line
    \deltadelta \lag / (\del_\t \psi_m) /   & = &
      - 2\i \eps^{mnp} \bpsi_n \sig_p = \bpi^m ,
\eeql
leading to the Poisson brackets
\beq
   \pois{\ee am}{\A_{nb}} = \i \eta_{ab} \delta_n^m, \ \ \ \
   \pois{\psi_m}{\bpi^n} = -\delta_m^n \, \eins.
\eeq
The constraints are the same as before, except that now
the Lorentz generator comes out as the derivative of $\lag$
with respect to the Lagrange multipliers $\A_{\t a}$:
\beql
  \H_A &=& - \i \eps^{mnp} \E mB \J_{aAB} \F_{npa}
            - 2\i \eps^{mnp} \, \bpsi_m \sig_A D_n \psi_p , \line
  \L_a & = & \i D_m \ee am + \ft\i2 \bpi^m \sig_a \psi_m , \line
  \S & = &  2\i \eps^{mnp} \sig_m D_n \psi_p , \line
  \bS & = &  -  D_m  \bpi^m .
\lab{con-const}
\eeql

In Ashtekar's representation the variables $\A_{ma}$ should appear
as multiplication operators and $\ee am$ is replaced by a differential
operator. To quantize the constraints it is therefore necessary
to write them as polynomials in $\ee am$. $\L'_a=\L_a$ and $\bS'=\bS$
                                             are
polynomials already, so only $\S$ has to be multiplied by a suitable
function of $\E nA$ to obtain a polynomial $\S'$.
The hamiltonian constraint is then obtained
by taking the bracket of $\bS'$ with $\S'$, which,
of course, gives again a polynomial.
It turns out that we have to define
\beq
  \S'  =  \ft12 \sig^C \, E\sig^\t \, \sig_C \, \S ,
\lab{S=ES'}
\eeq
where $E\sig^\t$ is the `densitized upper $\t$
component' of the curved Pauli matrices, which depends on the
`lower spatial components' $\E mA$ of the vierbein only.
Explicitly we have
\beq
   E\sig^\t = \ft16 \eps^{mnp} \eps_{ABCD} \, \sig^A \,
       \E mB \E nC \E pD.
\eeq
Note that this is the same transformation which
was found to make the 2+1 dimensional matter coupled
supersymmetry constraint in \cite{matschull.nicolai:93} a polynomial
in the canonically conjugate variables.

With the help of the polynomial $\S'$ we now define a new combination
of the hamiltonian and diffeomorphism constraints as
a matrix $\K'=\pois{\S'}{\bS'}$.
It is then equal to $\H_A\sig^A$, multiplied
by the matrix in \eq{S=ES'}, and up to a term proportional to
$\S$. So the complete set of constraints is
\beql
   \K' &=& \ft12 \i \F_{mna} \, \tsig^m \tsig^n \sig_a
          - 2\i \eps_{abc} \, \sig_a D_{[m} \psi_{n]}
             \, \ee bm \, \bpi^n \sig_c , \line
   \L'_a & =& \i D_m \ee am + \ft\i2 \bpi^m \sig_a \psi_m , \line
   \S' & =& - 2 \tsig^m \tsig^n D_{[m} \psi_{n]} , \line
   \bS' & =& - D_m \bpi^m ,
\lab{const'}
\eeql
where $\tsig^m=\ee am\sig_a$. Note that this is not the
four dimensional curved Pauli matrix with upper index.

Again, to solve the quantized constraints, we only need to solve
$\L'_a$, $\S'$ and $\bS'$. Choosing the representation
\beq
    \op{\ee am} = \hbar \, \deltadelta / \A_{ma} /,\ \ \ \
    \op{\bpi^p} = \i \hbar \, \deltadelta / \psi_m / ,
\eeq
they become exactly those given in \cite{jacobson:88}:
\beql
   \op{\L'_a} & = &
      \i\hbar D_m \Big( \deltadelta / \A_{ma} / \Big)
     + \ft12 \hbar \, (\sig_a \psi_m)_\alpha \,
      \deltadelta / \psi_{m\alpha}/, \line
   \op{\S'} & = &
       - 2\i \hbar^2 \eps_{abc} \, \sig_c D_m \psi_n
       \, \deltadelta / \A_{ma}/ \, \deltadelta / \A_{nb}/ , \line
   \op{\bS'} & = & - \i\hbar D_m \Big( \deltadelta / \psi_m / \Big ) .
\lab{susy-const}
\eeql

Given a wave functional $\Psi[\A_{ma},\psi_m]$, the constraints
$\L'_a$ and $\bS'$ just generate local Lorentz and supersymmetry
transformations, respectively. Thus they require $\Psi$ to be invariant
under $\delta\A_{ma}=D_m\lopar_a$, $\delta\psi_m=\ft\i2\lopar^a\sig_a%
\psi_m$ and under  the chiral supersymmetry transformation
$\delta\psi_m=D_m\epsilon$, which does not act on $\A_{ma}$.

In contrast to the metric representation, there are now purely
bosonic solutions to the constraints. In fact, they are just
{\em the same} formal solutions as those found in
               \cite{jacobson.smolin:88}
for bosonic gravity, namely the Wilson loop functionals
\beq
  T_\loop = \Tr \Pexp \ft\i2 \oint_\loop \d s \, \dot \loop^m(s)
   \, \A_{ma}(\loop(s)) \sig_a
\eeq
where $\loop(s)$ is a smooth, non-intersecting loop in the spatial
hypersurface, and $\dot\loop^m(s)$ denotes its tangent vector.
As is well known, when computing the second derivative of $T_\loop$
with respect to $\A_{ma}$ and $\A_{nb}$ one obtains an expression
symmetric in $m,n$. Thus $T_\loop$ is a solution to all the
constraints, because the second derivative in $\S'$ is antisymmetrized,
and $T_\loop$ also solves $\L'_a$ and $\bS'$, because it is
Lorentz invariant and does not depend on $\psi_m$.

There are some questions arising here. The first is: Where do these
purely bosonic solutions come from? They were not present in the
metric formalism. As was realized shortly after the discovery
of the loop solutions for pure gravity, they are annihilated
by the determinant of the spatial metric and therefore somehow
represent states with singular metric \cite{bruegmann.pullin:91}.
So far this is not a
serious problem, as Ashtekar's action and the constraints are
polynomials in the canonical variables and are thus able to handle
singular metrics.

But if the solutions represent states with singular metric, another
problem arises: As we multiplied the constraints by $E\sig^t$, which
is a third order polynomial of the inverse vierbein $\E mA$,
it seems that these solutions are not solutions to the
original constraints, but they are simply annihilated by this factor.
Note that the factors of $\ee am$ are ordered {\em to the right} in
$\S'$, thus the extra factors act on the wave functional first.

A second problem is that these solutions also solve the
constraints for pure gravity. If we drop all fermions from the
constraints, we are left with $\L'_a$ and $\K'$ only, and
$\K'$ again is proportional to the antisymmetrized second derivative
with respect to $\A_{ma}$. The solutions do not `see' the factor
$F_{mna}$ in $\K'$, because they are already annihilated by
$\ee a{[m}\ee b{n]}$. Thus we can replace the curvature
by the Rarita Schwinger field strength
$D_{[m}\psi_{n]}$ without changing the solutions,
and this replaces $\K'$ by $\S'$, thus switches from pure gravity
to supergravity.

This problem is somehow related to that arising when constructing
loop solutions for matter coupled gravity or when adding a cosmological
constant; e.g.~there are
extended Wilson loops solving the \WDW constraint of
gravity coupled to scalar fields \cite{matschull:93}, which do not
`see' the mass of these fields, nor do the loop states depend on the
cosmological constant.
It seems that they are just artificially
 generated by the extra factors
of the vierbein (or the dreibein in the gauge fixed version
often used) and do not correspond to solutions of the
\WDW equation in its original metric representation.

Obviously, this problem is closely related to the question how
to treat singular metrics, because multiplication of an equation
by a term that may become zero can change its solutions
drastically.
We will come to this problem again after discussing the three
dimensional theory

There is still another feature of the loop solutions for
supergravity. In the usual
connection representation of pure gravity, the loops are annihilated
by the
\WDW constraint but not by the diffeomorphism constraints, simply
because they are not invariant under spatial diffeomorphism.
One usually solves this problem by defining a loop representation
and requiring the wave functional depending on the
knot class of the loop only, which then is invariant under
diffeomorphisms.

But for supergravity we saw that $T_\loop$ solves all constraints,
though it is not invariant under diffeomorphisms.
To make this problem more explicit, observe that the
\WDW and diffeomorphism
constraints are both obtained from $\H_A$: The
diffeomorphism constraint is given by $\D'_m=\E mA\H_A$ and
the \WDW constraint is $\W'=EE^{\t A}\H_A$.

On the other hand, if we are allowed to obtain the \WDW operator
by multiplying $\H_A$ by a third order function of the vierbein,
there is no reason why this should not be allowed for the
diffeomorphism constraint. This is exactly what we have done
when using \eq{susy-const} as the
constraint algebra for supergravity.
Doing the same for pure gravity, the loop functionals
become solutions to {\em all} constraints, too: there
is no need to consider functionals on knot classes only in the
loop representation, as every arbitrary function which has
support on smooth loops only is a solution.

We will discuss all these problems in more detail in the last section.

\section{N=2 supergravity in three dimensions}

Let us now discuss the three dimensional theory, because some
of the problems concerning the loop solutions can be made more
explicit here as the theory can be solved exactly.
We use the same notation as in \cite{dewit.matschull.nicolai:93}.
The bosonic variables for N=2 supergravity are the
dreibein $\e \mu a$, $a=0,1,2$ the flat and $\mu=\t,\x,\y$ the curved
index, and the spin connection $\A_{\mu a}$ with curvature
\beq
   \F_{\mu\nu a} = \del_\mu \A_{\nu a} - \del_\nu \A_{\mu a}
      - \eps_{abc} \A_\mu\^b \A_\nu\^c ,
\eeq
where the flat indices are raised by the lorentzian metric
$\eta_{ab}={\rm diag}(-,+,+)$, and $\eps^{012}=-\eps_{012}=1$.
The gravitinos are again represented by 2-component complex
spinors $\psi_\mu$. As the Lorentz group $\grp SO(2,1)$ has a
real spinor representation $\grp SL(2,\R)$, we choose
real gamma matrices
\beq
   \gam_0 = \pmatrix{ 0 & 1 \cr -1 & 0 } , \ \ \ \
   \gam_1 = \pmatrix{ 0 & 1 \cr  1 & 0 } , \ \ \ \
   \gam_2 = \pmatrix{ 1 & 0 \cr 0 & -1 } ,
\eeq
obeying
\beq
   \gam_a \gam_b  = \eta_{ab}\eins - \eps_{abc} \gam^c.
\eeq
The Lorentz covariant derivative of a vector $v_a$, a
spinor $\chi$ and its conjugate $\bar\chi=\chi^\dagger(\i\gam_0)$
reads
\beql
       D_\mu v_a &=& \del_\mu v_a - \eps_{abc} \A_\mu\^b v^c , \line
       D_\mu \chi & = & \del_\mu \chi
                     + \ft12 \A_\mu\^a \, \gam_a \chi , \line
       D_\mu \bar\chi & = & \del_\mu \bar\chi
                     - \ft12 \A_\mu\^a \, \bar\chi \gam_a.
\eeql
The lagrangian is
\beq
  \lag = \ft12 \eps^{\mu\nu\rho} \e \mu a \F_{\nu\rho a}
     + 2 \eps^{\mu\nu\rho} \, \bpsi_\mu D_\nu \psi_\rho.
\lab{3-lag}
\eeq

Splitting space time into space and time, i.e.~splitting the
index $\mu$ into $\t$ and $i=\x,\y$, the configuration
variables become $\A_{ia}$ and $\psi_i$ with their momenta
\beq
   \deltadelta \lag / (\del_\t \A_{ia}) / =
         \eps^{ij} \e ja , \ \   \ \
   \deltadelta \lag / (\del_\t \psi_i) / =
     - 2 \eps^{ij} \bpsi_j.
\eeq
where $\eps^{\x\y}=\eps_{\x\y}=1$.
The Poisson brackets read
\beq
   \pois{\A_{ia}}{\e jb}= \eps_{ij} \delta_a^b, \ \ \ \
   \pois{\psi_i}{\bpsi_j} = \ft12 \eps_{ij} \, \eins.
\eeq
The lagrange multipliers $\e\t a$, $\A_{\t a}$, $\bpsi_\t$ and
$\psi_\t$ generate the constraints
\beql
   \H_a &=& \ft12 \eps^{ij} \F_{ija} , \line
   \L_a &=& \eps^{ij} D_i e_{ja} - \eps^{ij} \, \bpsi_i\gam_a\psi_j,
                                                 \line
   \S & = & 2 \eps^{ij} \, D_i \psi_j, \line
   \bS & =& 2 \eps^{ij} \, D_i \bpsi_j .
\lab{3-const}
\eeql
Note that their structure is similar to \eq{con-const}, but
they are much simpler, e.g.~only $\L_a$ contains the dreibein.
In particular, there are no operator ordering ambiguities and
thus no anomalies in the algebra. In quantum theory it is sufficient
to solve $\L_a$, $\S$ and $\bS$, as $\pois\S\bS=-\H_a\gam^a$.

\subsection*{The connection representation}
Here we briefly review the main
results of \cite{dewit.matschull.nicolai:93}.
The operators in the connection representation are
\beq
   \op{\e ia} = - \i\hbar  \eps_{ij} \deltadelta / \A_{ja} / , \ \ \ \
   \op{{\bpsi}}_{i\alpha}
   = \ft\i2 \hbar \eps_{ij} \deltadelta / \psi_{j\alpha} / .
\eeq
The constraints split into a set of multiplication operators
$\op{\H_a}$ and $\op\S$, and first order differential operators
$\op{\L_a}$ and $\op\bS$, which generate local Lorentz and
chiral supersymmetry transformations ($\delta\psi_i=D_i\epsilon$)
on the wave functional $\Psi[\A_{ia},\psi_i]$.

The solution to the constraints is given as follows.
$\op{\H_a}\Psi=0$ together with $\op\S\Psi=0$ implies that
$\Psi$ has support only on those fields $\A_{ia},\psi_i$
which have vanishing curvature $\F_{ija}=0$ and vanishing
supercurvature $D_{[i}\psi_{j]}=0$.
To simplify the notation we introduce a matrix valued connection field
$\A_i=\ft12\A_{ia}\gam^a$.
The covariant derivative of a spinor then reads
$D_i\phi=\del_i\phi+\A_i\phi$.

The complete set of curvature free pairs $\A_i,\psi_i$ is parametrized
by a $\grp SL(2,\R)$ matrix field $g$ and a spinor filed $\phi$ on
the covering manifold of the spatial 2-surface, which are subject to
certain relations between their values on points mapped to the
same point of the 2-surface
(see \cite{dewit.matschull.nicolai:93}).
These conditions are in fact equivalent to the requirement that
\beq
 \A_i[g]=g^{-1}\del_ig, \ \ \ \ \psi_i[g,\phi]=g^{-1}\del_i\phi
\eeq
are single valued on the 2-surface. Obviously, $\A_i[g]$ and
$\psi_i[g,\phi]$ are curvature free.

A complete set of solutions to $\op{\H_a}\Psi=0$ and $\op\S\Psi=0$
now reads
\beq
  \Phi_{g,\phi}[\A_i,\psi_i] =
   \prod_x  \delta\big(g^{-1}\del_ig -  \A_i \big) \,
      \prod_{x} \delta\big(  \del_i\phi - g \psi_i \big).
\lab{delta-sol}
\eeq
This is in fact an overcomplete set as e.g.~$g$ and $g_0g$, where
$g_0$ is a constant group element, or $\phi$ and $\phi+\phi_0$
with constant $\phi_0$, produce the same wave functional.
Nevertheless, a wave function can now be written as
\beq
  \Psi_F[\A_i,\psi_i] = \int \dd g \dd \phi \,
      F[g,\phi] \,  \Phi_{g,\phi}[\A_i,\psi_i],
\eeq
where the functional integral runs over all fields $g$ and $\phi$
subject to the restrictions mentioned above, and
$F[g,\phi]$ is an arbitrary functional.
The measure we assume to be invariant under multiplication of
$g$ with an arbitrary matrix field $g\mapsto gh$ and
under addition of a spinor field $\phi\mapsto\phi+\chi$.

The constraints $\op{\L_a}$ and $\op\bS$ now impose
additional conditions on $F$: $\op{\L}\Psi=0$ requires that
$\Psi$ is invariant under Lorentz transformations.
As can be verified, the change of $\Phi_{g,\phi}$ under
a Lorentz transformation with parameter $\lopar=\ft12\lopar^a\gam_a$,
i.e.
\beq
    \delta\A_i = \del_i \lopar-  [\lopar,\A_i] ,      \ \ \ \
    \delta\psi_i = - \lopar \psi_i ,
\eeq
can be compensated by changing $g$ as $\delta g=g\lopar$.
Similarly, a supersymmetry transformation generated by $\bS$ with
parameter $\epsilon$ is given by
\beq
    \delta \psi_i = D_i \epsilon,
\eeq
and the variation of $\Phi_{g,\phi}$ can be compensated by
$\delta\phi=g\epsilon$. So
we have to require that $F[g,\phi]$ is invariant under
Lorentz and supersymmetry transformations
\beq
  \delta g =  g \lopar , \ \ \ \
  \delta\phi= g\epsilon
\lab{F-inv}
\eeq
to obtain the full solution to all constraints.

For a 2-surface which is homeomorphic to $\R^2$ these restrictions
imply the $F$ is constant, as every field configuration
can be transformed into any other by \eq{F-inv}.
But remember that $\lopar$ and $\epsilon$ are single
valued fields on the 2-surface whereas $g$ and $\phi$
are fields on the covering manifold.
For nontrivial topologies there remain finitely many
degrees of freedom for $F$, called the moduli, which cannot be
gauged away by \eq{F-inv}. For a
more detailed discussion we again refer to
\cite{dewit.matschull.nicolai:93} and references therein.

\subsection*{The metric representation}
We will now show that the complete solution
to all the constraints can also be given in the metric
representation, and that it is equivalent to the connection
representation as it should be, because it is a quantum theory
with finitely many degrees of freedom and thus all representations
should be equivalent.

The transformation to the metric (or dreibein) representation
is obtained in two steps. First we introduce a `pseudo' metric
representation, which is simply the Fourrier transform of the
connection representation. As in the four dimensional theory
this is not the usual metric representation, because of an extra
multiplicator term in the Lorentz constraint. The
original metric representation is then obtained by an additional
canonical transformation.

In the pseudo metric representation the operators are just
given by interchanging the multiplication and differentiation
operators. The wave functional now depends on $\e ia$ and
$\bpsi_i$, the operators are
\beq
  \op{\A_{ia}}= -\i\hbar \eps_{ij} \deltadelta / \e ja / , \ \ \ \
  \op{\psi_{i\alpha}} =
   - \ft\i2 \hbar \eps_{ij} \deltadelta / \bpsi_{j\alpha} / ,
\eeq
and the constraints read
\beql
  \op{\H_a} &=& \eps^{ij} \del_i \op{\A_{ja}}
         - \ft12 \eps^{ij} \eps_{abc} \op{\A_i\^b} \op{\A_j\^c},\line
  \op{\L_a} &=& \eps^{ij} \del_i e_{ja}
        -   \eps^{ij} \eps_{abc} \e ib \op{\A_j\^c}
        - \eps^{ij} \bpsi_i \gam_a  \op{\psi_j} ,\line
  \op{\S} & = & 2\eps^{ij}\big(  \del_i  \op{\psi_j}
            + \op{\A_i} \, \op{\psi_j} \big), \line
  \op{\bS}  & = & 2\eps^{ij} \big( \del_i \bpsi_j
        - \bpsi_j \, \op{\A_i} \big).
\eeql
The complete solutions to these differential equations
can now be given by Fourrier transforming the solutions for the
connection representation. This procedure is rather simple
because we just have to replace the $\delta$-functions
in $\Phi_{g,\phi}$ by the corresponding exponentials.
We define
\beq
  \Phi_{g,\phi}[\e ia,\bpsi_i] =
    \exp \Big( \frac\i\hbar \intd2x \,
          \eps^{ij} \Tr(g^{-1}\del_i g \gam_j)
    -    2 \eps^{ij} \bpsi_i g^{-1} \del_j \phi  \Big).
\eeq
Then, using
\beq
   \deltadelta \Phi_{g,\phi} / \e ia / =
     - \frac\i\hbar \eps^{ij} \Tr(g^{-1}\del_jg \gam_a)
    \, \Phi_{g,\phi} \ \ \  \Rightarrow \ \ \
   \op{\A_i} \, \Phi_{g,\phi} =  g^{-1} \del_i g \, \Phi_{g,\phi}
\eeq
and
\beq
   \deltadelta \Phi_{g,\phi} / \bpsi_j / =
     -  \frac{2\i}\hbar \eps^{ij} g^{-1} \del_j \phi
    \, \Phi_{g,\phi} \ \ \  \Rightarrow \ \ \
   \op{\psi_i} \, \Phi_{g,\phi} =  g^{-1} \del_i \phi \, \Phi_{g,\Phi},
\eeq
one immediately finds that $\op{\H_a}\Phi=0$ and $\op{\S}\Phi=0$.
A straightforward calculation then shows that
acting on $\Phi$ with $\op{\L_a}$ or $\op{\bS}$ again gives
a variation that can be absorbed by a suitable transformation
of the type \eq{F-inv}.
As in the connection representation we define the state functional
by
\beq
  \Psi_F[\e ia,\bpsi_i] = \int \dd g \dd \phi \,
      F[g,\phi] \,  \Phi_{g,\phi}[\e ia,\bpsi_i],
\lab{PsiF}
\eeq
where $F$ again is a functional that is invariant under
the transformations \eq{F-inv}. Thus every state in the
connection representation corresponds to an equivalent
state in the pseudo metric representation and vice versa.

Observe that the states are constructed such that the constraints,
though they are second order differential operators, do not need
any regularization. Of course, the regularization is
`hidden' in the definition of $\Psi$ as a functional
integral over the fields $g$ and $\phi$, whose measure
has to be regulated somehow.

The final step from this pseudo metric to the usual dreibein
representation is similar to the procedure used in
the four dimensional theory to obtain a
Lorentz constraint that generates proper Lorentz
transformations on the wave functional.
Note that though $\Psi_F$ is a annihilated by $\op{\L_a}$,
is is not invariant under Lorentz transformations.

To obtain an invariant state functional,
we have to use another operator for the connection, i.e.
\beq
   \op{\A_{ia}} = - \i \hbar\eps_{ij}\,\deltadelta / \e ja /
             +    \eps_{ij} \, \deltadelta G[e] / \e ja /  .
\eeq
Again, $G$ has to be chosen such that its contribution to
$\op{\L_a}$ cancels against the derivative of the dreibein:
\beq
    \eps_{abc} \e ib \deltadelta G[e] / e_{ic} / =
        \eps^{ij} \del_i e_{ja} .
\eeq
A solution to this equation is
\beq
         G[e] = \ft12 \intd3x \eps^{ij} g^{kl} \eps_{abc}
                              \, \e ia \e kb \, \del_j \e lc ,
\eeq
where $g^{kl}$ is the inverse of the two dimensional metric
$g_{kl}=\e la e_{ka}$.
The derivative of $G$ with respect to the dreibein will be called
$\omega_{ia}$. Note that this can not be interpreted as the spin
connection for which the dreibein becomes covariantly constant:
we are dealing here with a {\em drei\/}bein on a {\em two} dimensional
surface, i.e.~it is not invertable and the requirement that
the torsion $\del_{[i}e_{j]a}-\eps_{abc}\omega_{[i}\^b\e {j]}c$
vanishes
does not imply that the full covariant derivative of $\e ia$
vanishes.
The new operator for $\A_{ia}$ now becomes
\beq
     \op{\A_{ia}} = \omega_{ia}
                    -\i\hbar \eps_{ij} \, \deltadelta / \e ja / .
\eeq
Inserting this into the constraints
we obtain
\beql
    \op{\L_a}& =&
     - \i \hbar \eps_{abc} \e ib \, \deltadelta / e_{ic} /
     - \ft12 \hbar \bpsi_i \gam_a \, \deltadelta / \bpsi_i / ,
              \line
    \op{\S} & = &
     \i \hbar \nabl_i \Big (\deltadelta / \bpsi_i / \Big)
     -\ft12 \hbar^2 \eps_{ij} \gam_a \, \deltadelta / \bpsi_i /
                         \, \deltadelta / \e ja / , \line
    \op{\bS} & = &
      2 \eps^{ij} \, \nabl_i \bpsi_j + \i \hbar \,
             \bpsi_i \gam^a \, \deltadelta / \e ia / ,
\lab{3metric-const}
\eeql
where $\nabl_i$ again denotes the covariant derivative with respect to
$\omega_{ia}$, i.e.
\beq
   \nabl_{[i} \bpsi_{j]} = \del_{[i} \bpsi_{j]}
     - \ft12 \omega_{[ia} \, \bpsi_{j]} \gam^a
\eeq
For the \WDW operator we get a slightly cumbersome expression
which we will not give explicitly here: it is still the commutator
of $\op{\bS}$ with $\op{\S}$.
The constraints are now formally the same as those for the
four dimensional theory, except that they do not contain the
inverse vierbein

The complete solutions to these constraints are now given by
multiplying the old solutions with the exponential of the functional
$G$, i.e.~replacing $\Phi$ by
\beq
    \widetilde\Phi_{g,\phi}[\e ia,\bpsi_i] =
    \exp\Big(-\frac\i\hbar G[\e ia] \Big) \,
              \Phi_{g,\phi}[\e ia,\bpsi_i].
\eeq

Inserting $\widetilde\Phi$ instead of $\Phi$ into \eq{PsiF}
then gives the solution to the constraints in the
metric representation.

\subsection{The densitized connection representation}
The connection representation for the three dimensional theory
discussed above is not directly related to that of the
four dimensional theory. In particular, because of the simple
structure of the constraints, it is not necessary to multiply
the constraints by extra factors of the dreibein to obtain
polynomial expressions.

The situation is different when we consider, e.g., the dimensionally
reduced version of the four dimensional N=1 theory, which
contains additional matter fields, but whose graviton and gravitino
action is the same as that of the N=2 theory discussed here.
In \cite{matschull.nicolai:93} we found that for the matter coupled
theory we also have to multiply one of the supersymmetry constraints
by an extra factor of $e\gam^\t=-\ft12\eps^{ij}\eps_{abc}\gam^a\e ib
\e jc$ to get polynomial constraints,
which directly corresponds to the definition
\eq{S=ES'}.

Thus the question arising here is what happens if we start with
the constraints given in \cite{matschull.nicolai:93} for the
matter coupled theory and just drop all the matter terms.
We then obtain
\beql
   \L'_a & = &  \eps^{ij} D_i e_{ja} - \eps^{ij}\bpsi_i\gam_a\psi_j
                ,  \line
   \S' & = & - \eps^{ij}\eps^{kl}
                         \eps_{abc} \e ka \e lb \, \gam^c D_i \psi_j
           , \line
   \bS' & = & 2 \eps^{ij} D_i \bpsi_j .
\eeql
They are just the three dimensional versions of \eq{const'} and
can be constructed in the same way, i.e.~multiplying $\S$ by
$e\gam^\t$ and then defining $\H'_a$ as the bracket of
$\bS'$ with $\S'$. A priori there is no reason why these
constraints could not be regarded as the canonical constraints
of 2+1 supergravity, if \eq{const'} describes 3+1 supergravity.
They are nothing but their three dimensional counterparts, and
assumed we didn't know that there is a much simpler expression
for them, we had to use them to quantize the 2+1 theory.

The quantized versions
of the Lorentz and supersymmetry constraints now become
\beql
   \op{\L'_a} & = &
        \i \hbar D_i \Big( \deltadelta / \A_{ia} / \Big)
      -  \ft\i2 \hbar\,  (\gam_a \psi_i )_{\alpha}
         \, \deltadelta / \psi_{i\alpha} / , \line
  \op{\S'} & = &
    -  \hbar^2 \eps^{ij}\eps_{abc} \, \gam^c D_i \psi_j
      \, \deltadelta / A_{ia} / \, \deltadelta / A_{jb} / , \line
  \op{\bS'} & = &
    - \i \hbar \, D_i \Big( \deltadelta / \psi_i / \Big) .
\eeql

Their structure is again similar to \eq{susy-const} and, in fact,
we can also give the Wilson loop solutions to them:
\beq
  T_\loop = \Tr \Pexp \ft12 \oint_\loop \d s \, \dot \loop^i(s)
   \, \A_{ia}(\loop(s)) \gam^a,
\eeq
which is obtained from the 3+1 version just by replacing the
$\sig_a$-matrices by $\gam_a$. Again, $\bS'$ and $\L'_a$ are
solved by $T_\loop$ because it is Lorentz invariant and does not
depend on the gravitino. If the loop is smooth and does not
have self intersections, the remaining constraints are solved
because of the same symmetry arguments as in 3+1 dimensions,
thus every smooth loop embedded in the spatial surface yields
a state functional.

This is a completely different result than that obtained
above, where we found that the state is described by a function
$F$ that depends on finitely many degrees of freedom, the moduli,
only. Here the general solution is given as an arbitrary
superposition of $T_\loop$'s.
In particular, the constraints do not imply that the
curvature vanishes and thus different loops really produce
different wave functionals.

At first sight, the problem seems to be that we are
dealing with single loops here instead of equivalence
classes of loops under diffeomorphisms. We already mentioned this
problem for the four dimensional theory: it arises because
we also multiplied the diffeomorphism constraint by a dreibein
factor. The usual way to obtain the diffeomorphism
and \WDW constraint from $\H_a$ would in this case be
$\D'_i=\e ia \H_a$ and $\W'=\ft12\eps_{abc}\eps^{ij}\e ia\e jb\H^c$.

Then $\D'_i$ in fact generates diffeomorphisms on the 2-surface
and requires $\Psi$ to be invariant, and $\W'$ is automatically
solved for every solution of $\S'$ and $\bS'$.
Doing this, however, the result is still totally different.
Using the loop representation, the wave function becomes
a function on the generalized knot classes of the
2-surface. But there are still infinitly many\footnote
{Of course, there are no real `knots' on a 2-surface,
but there are infinitly many link classes, given e.g.~by
different numbers of disconnected loops.}, even if the
topology is trivial, whereas for the metric or `undensitized'
connection representation there is only one state
$\Psi_F$ with constant $F$.

\section{Discussion}

In the 2+1 case described above it is rather obvious that
the loop solutions are artificially generated by multiplying
the constraints with a factor, which is in principle the
determinant of the spatial metric. As this factor appears
to the right in the quantized constraints and already
annihilates the loop functionals, it is not clear
what they have to do with the original theory described
by \eq{3-const}. Here we cannot argue that the
`polynomialized' primed constraints can deal with singular
metrics and thus we get new solutions representing
singular metrics: the constraints were polynomial from the
beginning and the unprimed constraints can handle singular
metrics, too.

To make this argument more precise, consider the
classical theory described by the primed and unprimed constraints.
If we allow the dreibein to be singular, there are solutions
to the primed constraints which do not solve the unprimed ones.
Just take the trivial example $\e ia=0,\psi_i=0$, but
$\A_{ia}$ not curvature free.

Thus the primed constraints do not
describe the classical theory defined by the lagrangian
\eq{3-lag}, if we allow singular
metrics, though the lagrangian itself does not contain
the inverse metric.
A lagrangian which directly leads to the primed constraint
can be written in the same way as one usually writes the
Einstein Hilbert lagrangian in Ashtekars variables with the
time component of the dreibein split into a lapse and shift
function. One has to replace the Lagrange multipliers $\e\t a$
by $\e\t a=n^i\e ia+n\eps^{abc}\eps^{ij}e_{ib}e_{jc}$, and a similar
replacement for $\bpsi_t$, but let us drop the fermions
now because the following arguments also apply for pure gravity.

Differentiation with respect to $n$ and $n^i$ then directly
yields the primed diffeomorphism and hamiltonian constraints.
But the action
is no longer invariant under space time diffeomorphisms.
The transformation of the variables $n$, $n^i$ and $\e ia$
under the full 2+1 dimensional diffeomorphisms involve
the inverse of the metric determinant, as $n$ is a density of
weight $-1$.

The argument can be transfered to the four dimensional theory as well.
The lagrangian in term of Ashtekar's variables we started from
(see \eq{lag-asht}) was polynomial in $\E MA$, thus it is well
defined for singular metrics.\footnote{Observe that also the
real first order Einstein Hilbert action is polynomial. But
when treating it canonically there arise
second class constraints making the dirac brackets non-polynomial.
It is this fact and not the polynomial action that simplifies
the canonical treatment with Ashtekar's variables.}
Again, to obtain the primed constraints directly,
one has to introduce a lapse and shift function by
\beq
   \E \t A = N^m \E mA - \ft16 N \eps^A\_{BCD} \eps^{mnp}
                          \E mB \E nC \E pD .
\lab{4-lapse}
\eeq
Thus there are two different actions for 3+1 gravity, one
given by \eq{lag-asht} and the other one by inserting
\eq{4-lapse} and using $N,N^m$ instead of $\E \t A$ as
the primary field variables. Both actions induce the same
equations of motion for non-singular metrics, and both
are well defined for (different kinds of) singular metrics.

As in the 2+1 theory
the equations of motion are different for the different actions
if the metric becomes singular.
It seems that there is a priori no reason why one lagrangian
can be better than the other, because we do not know the
`correct' equations of motion. But there {\em is} a crucial
difference between the two lagrangians: Only one of them
is invariant under the full diffeomorphism group of space time.
If one requires
the invariance of general relativity under all
diffeomorphisms {\em and} allows singular metrics, one
has to use the action \eq{lag-asht}.
We can conclude that out of the following three
properties of classical canonical gravity only two can be realized
simultaneously:
\begin{itemize}
\item[-] invariance under space time diffeomorphism,
\item[-] polynomial constraints,
\item[-] singular metrics allowed.
\end{itemize}

This classical argument is in agreement with our conclusion
concerning the loop solutions in the quantized theory. We saw that,
if we also `densitize' the diffeomorphism constraint, which came
out automatically in supergravity, then the loops became
solutions to all constraints without considering wave
functions that depend on the knot class only. Thus the
invariance under spatial diffeomorphisms is destroyed by
multiplying the corresponding constraint by extra factors
of the vierbein.
Now we saw that in the classical theory the invariance
under space time diffeomorphisms is lost when introducing
the lapse function $N$ of weight $-1$, i.e.~when going over
from the unprimed to the primed constraints.
Assuming that the arguments above apply to the quantized theory as
well, we must conclude that quantum gravity described by the
loop states is not invariant under space time diffeomorphisms,
because it explicitly needs the polynomial constraints and
singular metrics. This is  rather unsatisfactory,
because invariance under space time diffeomorphisms is
one of the first principles of general relativity: in fact, it
{\em is} `general relativity'.

So what is the conclusion out of this? Let us first note that
the problem does not arise with the use of Ashtekar's
variables. They can be eliminated by going back to the
real action with $\E MA$ and $\O_{MAB}$ as primary fields.
All the arguments concerning the classical theory apply to
this action as well, except that there is no `reason' to
introduce a densitized lapse function because this would not
simplify the canonical constraints considerable.

The problem arises when one has to give up one of the
three properties listed above. For practical reasons it is
most convenient
to drop general invariance, because in the
canonical formulation the {\em manifest} invariance is lost anyway.
But, as already mentioned, this should be the last principle
given up when dealing with general relativity.
If one insists on 3+1 diffeomorphism invariance and singular
metrics, then
one has to start with the unprimed constraints and one
is not allowed to multiply them by factors which may become
zero. In quantum theory it is even more inappropriate to
order these extra factors {\em to the right}.

On the other hand, if singular metrics are not allowed, then
the different actions and different sets of constraints are
equivalent. Nevertheless there is still a problem with
the operator ordering. Note here that it is
the ordering with the extra factors appearing {\em to the left}
in the \WDW operator which leads to a closed algebra of
constraints, whereas there is an anomaly in the algebra
admitting the loop solutions
(see \cite{bruegmann.gambini.pullin:92b} for a
discussion of different factor orderings).

This is
an interesting result, because the
`Chern Simons' state found in
\cite{bruegmann.gambini.pullin:92b} solves the constraints
in the opposite factor ordering (and with a cosmological constant)
and it does not correspond to a singular metric. In fact, one doesn't
have to
make use of the $\ee am$ factors in the \WDW constraints to
show that the Chern Simons form is a solution.
It really `solves' the non-polynomial constraints, where
the quotation marks shall indicate that the problem is
how to define the quantized versions of the non-polynomial
constraints properly.

These properties of the Chern Simons solution
are easily proofed. As they only exist for a non-vanishing
cosmological constant, we add to $\lag$
\beql
  \ft12 \CC   E  & = &
 -\ft1{48} \, \CC \, \eps^{MNPQ} \eps_{ABCD} \, \E MA \E NB \E PC \E QD
  \line  & = &
   -\ft\i{12} \, \CC \, \eps^{MNPQ} \J_{aAB} \J_{aCD}
          \E MA \E NB \E QC \E PD,
\eeql
where we used that $\J_{a[AB}\J_{aCD]}=-\ft\i4\eps_{ABCD}$.
Now the diffeomorphism and
hamiltonian constraints become
\beq
   \H_A = - \i  \E mB \J_{aAB}
  \big( \eps^{mnp} \F_{npa} -  \ft13 \CC \ee am \big) .
\lab{cosmo-ham}
\eeq
Obviously, this constraint has a simple solution in the
connection representation, namely the
exponential of the Chern Simons form, as its
derivative yields the field strength:
\beq
  \Psi = \exp \Big(\frac1{\hbar\CC} \intd3x
    \eps^{mnp} (  3\A_{ma}\del_n\A_{pa} -
                   \eps^{abc}\A_{ma}\A_{nb}\A_{pc} ) \, \Big)
\eeq
obeys
\beq
    \Big(  \eps^{mnp} \F_{npa}  -
          \ft13 \hbar\CC, \deltadelta / \A_m\^a / \Big) \Psi = 0 .
\eeq
Of course, there are some problems with this functional, too; e.g.~%
it is not clear whether it is a normalizable state functional.
But let us ignore these problems for the moment and discuss
only the properties related to the operator ordering.

We see that $\Psi$ is annihilated by $\op{\H_A}$, though this
is not yet defined properly, because it is not clear
how to represent $\E mA$ in the connection representation.
But the solution does not depend on how it is defined.

Thus the question arises whether one can make sense out of the
non-polynomial expression $\H_A$ in the connection representation.
A possible answer is the following.
The action of
$\op{C\H_A}$ on $\Psi$ should give zero for every function
$C[\E mA]$ that makes $C\H_A$ a polynomial in $\ee am$.
Of course, this only makes sense if the extra
factors always appear to the left of $\H_A$.
This would just lead to the polynomial \WDW and diffeomorphism
constraints with the dreibein factors ordered to the left,
as there are only four independent choices for $C$ that
make $C\H_A$ a polynomial. So
the Chern Simons functional is a solution of this type,
whereas the Wilson loops are not.

Maybe this procedure to treat a non-polynomial constraint
is as close as possible to the classical
theory which is invariant under the complete 3+1
diffeomorphism group. Thus there are more reasons to prefer
the operator ordering where the dreibein appears to the left than
just the fact that only then the quantum algebra closes.

Finally we should emphasize that, even if we do not use
the polynomialized constraints, Ashtekar's variables
still simplify the canonical treatment in a way that may have been
overlooked because the polynomial form of the constraints
seems to be their `simplest' form and only this has been
studied so far.

But there is still a difference between the constraint
$\H_A$ in terms of Ashtekar's variables given in
\eq{cosmo-ham} and the original \WDW operator constructed
in the metric representation \cite{wheeler:64,dewitt:67},
which contains the curvature scalar of the spatial
metric and thus the inverse metric.
However, the only non-polynomial term in
$\H_A$ is the vierbein component $\E mA$, which is
not the inverse but a kind of `square root' of the canonical variable
$\ee am$ (see the definition \eq{ee-def}), and
a square root of a differential operator is sometimes less
a problem then to define its inverse; a well know example is
the Dirac operator as `square root' of the Klein Gordon
operator.

So the question is whether it is possible to define an
operator $\op{\E mA}$ in the connection representation such that
\beq
    \komm{\op{\E mA}}{\op{\E nB}}=0, \ \ \ \
    \komm{\eps^{mnp} \op{\E mA} \op{\E nB} \J_{aAB}}{\A_{qb}} =
       - \hbar \delta^p_q \eta_{ab}.
\eeq
Then
$\H_A$ becomes well defined and the Chern Simons functional
would be a well defined solution to the constraints
which correspond to the space time diffeomorphism invariant
and singular metric allowing version of Einstein gravity.

\end{document}